\documentclass[12pt]{article}
\usepackage[totalwidth=460truept,totalheight=622truept]{geometry}
\usepackage{latexsym,graphicx,amsfonts,amssymb,amsmath,xcolor}
\usepackage[hypertexnames=false,hidelinks]{hyperref}
\usepackage{fancyhdr}
\usepackage{xcolor}

\pdfoutput=1

\def\theequation{\arabic{section}.\arabic{equation}}

\renewcommand{\theequation}{\thesection.\arabic{equation}}
\global\arraycolsep=1truept
\renewcommand{\theequation}{\arabic{section}.\arabic{equation}}

\begin{document}

\bigskip \phantom{C}

\vskip2truecm

\begin{center}
{\huge \textbf{Fractions and Fakeons }}

\vskip.5truecm

{\huge \textbf{in\ Quantum Field Theory}}

\vskip1truecm

\textsl{Damiano Anselmi}

\vskip.5truecm

\textit{Dipartimento di Fisica \textquotedblleft Enrico
Fermi\textquotedblright , Universit\`{a} di Pisa,}

\textit{Largo B. Pontecorvo 3, 56127 Pisa, Italy}

\textit{and INFN, Sezione di Pisa, Largo B. Pontecorvo 3, 56127 Pisa, Italy,}

damiano.anselmi@unipi.it

\vskip2truecm

\textbf{Abstract}
\end{center}

We investigate formulations of quantum field theories whose kinetic terms
involve fractional or continuous powers of the d'Alembert operator. The
primary requirements are perturbative unitarity and a well-defined classical
limit with a finite number of initial conditions. A direct approach consists
of continuing the correlation functions from Euclidean space to Minkowski
spacetime using the fakeon prescription for the fractional part of the
power. Alternative formulations arise through decomposition, in which the
fractional part is represented as a continuum of ordinary fakeons. These
options are infinite in number and yield inequivalent Minkowskian theories
with the same Euclidean counterpart. We demonstrate these features at tree
level and for bubble diagrams. We also point out potential pitfalls in the
calculations. Finally, we show how to treat continuous powers of covariant
d'Alembertians in fractional gauge and gravity theories. The Ward and
Cutkosky identities hold in all formulations.

\vfill\eject

\section{Introduction}

\label{intro}\setcounter{equation}{0}

Quantum field theory (QFT) remains by far the most successful framework for
the description of high-energy phenomena. It forms the basis of the Standard
Model of particle physics and has achieved an extraordinary level of
agreement with experiments. While a variety of alternative approaches to
fundamental interactions have been proposed in the past decades, none have
yet matched the empirical success and internal consistency of QFT.

Despite its accomplishments, extending QFT beyond its standard domain or
modifying its fundamental structures raises subtle conceptual and technical
challenges. In this context, theories characterized by nonlocal or
generalized kinetic operators have attracted considerable attention,
starting with the pioneering works of Pais and Uhlenbeck \cite{Pais} and
Efimov \cite{efimovnl}. Various developments have surfaced since then. Among
these we highlight the intriguing strategy pursued by Krasnikov, Kuz'min,
Tomboulis, Modesto and others over the years \cite%
{Krasnikov,kuzmin,Tomboulis,Modesto,Marcus,Lanza}, which amounts to removing
the ghost degrees of freedom of higher-derivative theories by inserting
appropriate form factors. In this way, perturbative unitarity is achieved 
\cite{efimov,briscese} and renormalizability is preserved \cite{Marcus,Lanza}%
. Related investigations worth mentioning are those of refs. \cite{Koshelev}.

Nevertheless, not everybody is comfortable with abandoning locality
altogether. A different strategy was outlined by Lee and Wick at the end of
the sixties \cite{leewick}, where the \textquotedblleft abnormal
particles\textquotedblright\ decay quickly enough to remain unobservable in
most applications. Variants of this idea have been proposed in the past
decade, involving propagators with complex poles \cite{related}, analogies
with Quantum Chromodynamics (QCD) \cite{Holdom}, antilinear symmetries \cite%
{Mannheim} and unstable ghosts \cite{Donoghue}.

A sort of intermediate option between locality and nonlocality is
represented by purely virtual particles, or \textquotedblleft
fakeons\textquotedblright , which\ are \textquotedblleft
particles\textquotedblright\ that are always off the mass shell \cite%
{LWfakeons}. One starts from a \textquotedblleft parent\textquotedblright\
local theory and generates the \textquotedblleft
descendant\textquotedblright\ nonlocal fakeon theory by turning would-be
ghosts or even non-problematic particles into fake particles through certain
diagrammatic operations and a projection on the asymptotic states \cite%
{Piva,LWfakeons,LWgrav,diagrammarMio,PVP20}. This procedure yields a
nonlocal quantum field theory of a special type, which can be renormalizable
and unitary at the same time.

Fakeons leave observable imprints of various types. In quantum gravity, they
lead to testable predictions about primordial cosmology by affecting the
spectra of scalar and tensor fluctuations \cite{ABP}. From the
phenomenological point of view, they differ from resonances by exhibiting a
pair of bumps rather than a peak \cite{peak}. Moreover, they trigger a
violation of microcausality \cite{causalityQG} through the breaking of time
ordering at energies larger than the fakeon mass \cite{PVP20}.

A further arena to explore departures from locality while preserving key
structural properties of QFT is fractional quantum field theory \cite%
{Calca,CalcaRach,Calcarest,Calcarest2}, which involves rational or real
powers of the d'Alembert operator. The main challenge posed by these models
is to have unitarity and a consistent classical limit at the same time.

Calcagni and Rachwa\l\ were the first to use fakeons as tools to address the
unitarity issue \cite{CalcaRach}, through spectral representations of
complex powers of the d'Alembert operator \cite{Calcarest}. In ref. \cite%
{Calcarest2} models of fractional quantum gravity were studied.

We also recall that models with continuum powers of kinetic operators,
defined by means of the usual Feynman prescription, have been considered by
Krasnikov \cite{kras} as particular cases of the notion of unparticle,
introduced by Georgi in ref. \cite{unpart}. In those models the issue of
unitarity is addressed from the point of view of conformal field theory.

Here we elaborate on fractional QFT in a general field theoretical spirit.
We prove that fakeons can be employed in infinitely many inequivalent ways,
starting from a \textquotedblleft direct\textquotedblright\ formulation,
which does not necessitate the use of spectral representations. Each option
yields a different Minkowskian theory with the same Euclidean counterpart.

In the direct\ approach the Minkowskian correlation functions are obtained
by applying the fakeon prescription (via the so-called \textquotedblleft
average continuation\textquotedblright\ \cite{Piva,LWfakeons,LWgrav})
directly to the Euclidean ones.

The other formulations consist of converting the fractional powers of
derivative operators into one-parameter families of ordinary fakeons through
spectral decompositions. We analyze options not considered in the literature
so far. A particular choice returns the results of the direct approach.

Depending on the decomposition itself, one obtains an infinite class of
inequivalent models. We illustrate this explicitly at tree level and in loop
diagrams, pointing out nontrivial aspects of the calculations, which may
lead to erroneous results if fakeons are not dealt with by the book.

Fakeons ensure perturbative unitarity, which is expressed diagrammatically
by the Cutkosky-Veltman identities \cite{cutkosky}. Another requirement is
that the theory admit a proper classical limit. This is not guaranteed, in
general, since fractional powers of the d'Alembert operator may lead to
non-real or non-Hermitian kinetic terms. All formulations studied here admit
a meaningful classical limit and propagate only the expected physical
degrees of freedom. In particular, the nonlocal field equations are not
burdened with the need to specify infinitely many initial conditions.

Finally, we show how to treat fractional or continuous powers of covariant
d'Alem\-ber\-ti\-ans, in kinetic terms and vertices, to prove that the
models can be consistently coupled to gauge fields and gravity. The Ward
identities are satisfied in all formulations.

The paper is organized as follows. In section \ref{frac} we outline various
options on the table to formulate fractional models, and show their
inequivalence at tree level. In section \ref{diagra} we calculate the bubble
diagram in the simplest models and show that different formulations yield
different results. In section \ref{extfrac} more general fractional models
are treated. Fractional powers of covariant d'Alembertians are handled in
section \ref{covfrac}. In section \ref{class} we prove that the set of
degrees of freedom is the expected one. In appendix \ref{A} we list simple
properties of noninteger powers of a complex variable, while in appendix \ref%
{B} we study the analytic continuation of the bubble diagram in the
decomposition approach.

Square roots, logarithms, and non-integer powers are all taken on the
principal branch. When a regularization technique is needed, we use the
dimensional one \cite{dimreg}, which aligns well with continuous powers of
kinetic operators. Seen as complex parameters, they are at the core of the
so-called analytic regularization \cite{anareg}, of which the dimensional
one is the manifestly gauge-invariant upgrade.

\section{Models with a fractional d'Alembertian}

\label{frac}\setcounter{equation}{0}

In this section we consider the simplest fractional models, based on the
kinetic operator $\Box ^{\gamma }$, comparing various options for their
formulations. We introduce the direct approach, which has not appeared in
the literature so far, and point out that it is not universally applicable.
When it applies, we show how to retrieve equivalent results within the
decomposition approach. Finally, we compare several QFT aspects of different
decomposition formulations.

Let us start from the classical Lagrangians%
\begin{equation}
\mathcal{L}_{1}=-\frac{1}{2}\phi \Box ^{\gamma }\phi -\frac{\lambda }{3!}%
\phi ^{3},\qquad \mathcal{L}_{2}=-\frac{1}{2}\phi \left( \Box ^{2}\right)
^{\gamma /2}\phi -\frac{\lambda }{3!}\phi ^{3},  \label{two}
\end{equation}%
and assume $\gamma <2$ to avoid infrared problems of the type $\sim \int 
\mathrm{d}^{D}p/(p^{2})^{\gamma }$ in loop integrals around $d=4$. We also
assume $\gamma >0$ to have propagators that tend to zero in the ultraviolet
limit.

The theories (\ref{two}) coincide in Euclidean space, since $\Box ^{\gamma
}=(-p_{\text{M}}^{2})^{\gamma }=(p_{\text{E}}^{2})^{\gamma }=\left[ (p_{%
\text{E}}^{2})^{2}\right] ^{\gamma /2}=\left[ (-p_{\text{M}}^{2})^{2}\right]
^{\gamma /2}=\left( \Box ^{2}\right) ^{\gamma /2}$ (where $p_{\text{M}}$ and 
$p_{\text{E}}$ are the Minkowski and Euclidean momenta, respectively), but
differ in Minkowski spacetime. A shortcoming of the first Lagrangian is that
it is not real for non-integer $\gamma $, at least if we interpret $\Box
^{\gamma }$ as it stands. The classical equations of motion are%
\begin{equation}
\Box ^{\gamma }\phi +\frac{\lambda }{2}\phi ^{2}=0,  \label{feq}
\end{equation}%
and the propagator $-i/z^{\gamma }$, with $z=-p^{2}$, involves a fractional
power that must be defined for $z\,<0$. If we choose the branch cut on the
negative real axis, the field equations are ill defined. If we choose it
somewhere else, they are not real. The good news is that the fakeon approach
is directly applicable to the Lagrangian $\mathcal{L}_{1}$ (see below).

The second theory leads to real classical equations of motion, 
\begin{equation}
\left( \Box ^{2}\right) ^{\gamma /2}\phi +\frac{\lambda }{2}\phi ^{2}=0.
\label{feq2}
\end{equation}%
Here we face a different type of problem. The complexification of the
propagator $-i/(z^{2})^{\gamma /2}$, consists of two analytic functions with
a cut along the imaginary axis. See Appendix \ref{A} for details. This means
that we cannot apply the fakeon approach as is, because it is based on the
average continuation of \textit{one} analytic function. Although
generalizations might exist, we do not pursue them here.

The inherent differences between the two models suggest that they may lead
to inequivalent theories in Minkowski spacetime, as we are going to show.

\subsection{Direct fakeon approach}

The direct fakeon approach is based on the \textquotedblleft average
continuation\textquotedblright\ \cite{Piva,LWfakeons,LWgrav} of the
Euclidean amplitudes to Minkowski spacetime, which amounts to averaging the
analytic continuations around the branch cuts involving fakeons. In the
models we are considering, which do not propagate degrees of freedom, all
the branch cuts are encompassed by the fakeon prescription.

This method is applicable to $\mathcal{L}_{1} $, but not to $\mathcal{L}_{2}$%
, because the propagator $-i/(z^{2})^{\gamma /2}$ breaks the complex plane
into two disjoint regions.

When we apply the average continuation to the $\mathcal{L}_{1}$ propagator,
encoded into the function $f(z)=-i/z^{\gamma }$, the classical limit of $%
\mathcal{L}_{1}$ becomes acceptable. At first sight, it may seem unusual to
advocate the average continuation already at tree level, but that is what
one also does in local, non-fractional theories \cite{Piva,LWfakeons,LWgrav}%
. There the operation is so simple that it basically goes unnoticed: acting
on simple poles instead of cuts, it returns the Cauchy principal value $%
\mathcal{P}$, which defines the propagator \textquotedblleft on the
pole\textquotedblright . For example, the average continuation of $1/p^{2}$
is $(1/2)[1/(p^{2}+i\epsilon )+1/(p^{2}-i\epsilon )]=\mathcal{P}(1/p^{2})$.

Writing $z=\mathrm{e}^{i\theta }$ and noting that on the negative real axis%
\begin{equation*}
\frac{1}{2}\lim_{\theta \rightarrow \pi }\mathrm{e}^{-i\gamma \theta }+\frac{%
1}{2}\lim_{\theta \rightarrow -\pi }\mathrm{e}^{-i\gamma \theta }=\cos
\left( \pi \gamma \right) ,
\end{equation*}%
the $\mathcal{L}_{1}$ average-continued propagator in momentum space is 
\begin{equation}
G_{1}(p^{2})=-\frac{i\theta (-p^{2})}{(-p^{2})^{\gamma }}-\cos \left( \pi
\gamma \right) \frac{i\theta (p^{2})}{(p^{2})^{\gamma }}=-i\hspace{0.01in}%
\text{Re}\left[ \frac{1}{(-p^{2}-i\epsilon )^{\gamma }}\right] .  \label{G1}
\end{equation}%
For reference, the fakeon Green function in coordinate space, equal to the
Fourier transform of $G_{1}(p^{2})$, reads%
\begin{equation*}
\tilde{G}_{1}(x^{2})=\int \frac{\mathrm{d}^{D}p}{(2\pi )^{D}}G_{1}(p^{2})%
\mathrm{e}^{-ip\cdot x}=-\frac{i\Gamma (\frac{D}{2}-\gamma )}{2^{2\gamma
}\pi ^{D/2}\Gamma (\gamma )}\text{Re}\left[ \frac{i}{(-x^{2}+i\epsilon
)^{D/2-\gamma }}\right] .
\end{equation*}%
This result can be easily checked in the limit $\gamma \rightarrow 1$.

The effective action%
\begin{equation}
\Gamma _{1}=-\frac{1}{2}\phi \left[ \theta (\Box )\Box ^{\gamma }+\sec
\left( \pi \gamma \right) \theta (-\Box )(-\Box )^{\gamma }\right] \phi -%
\frac{\lambda }{3!}\phi ^{3}+\text{ radiative corrections}  \label{Lc}
\end{equation}%
makes sense for $\gamma \neq 1/2$, $3/2$. For $\gamma $ semi-integer $\Gamma
_{1}$ is singular, so it is convenient to use the generating functional $%
W_{1}$ of connected Green functions, which is well defined.

At the level of radiative corrections we proceed similarly, by
average-continuing the Euclidean diagrams and the correlation functions to
Minkowski spacetime. At the practical level, it is convenient to use the
dimensional regularization \cite{dimreg}, where $D=d-\varepsilon $ denotes
the complex spacetime dimension and $d$ is the physical one. It is also
convenient to treat $\gamma $ as a complex parameter, in the spirit of the
analytic regularization \cite{anareg}. In this way, most loop integrals
become ultraviolet convergent.

Consider, for example, the bubble diagram. In $d=4$ dimensions the
associated integral is ultraviolet convergent for $1<\gamma <2$, and can be
extended analytically to $0<\gamma <1$. We can assume $\gamma \neq 1$ and
reach the case $\gamma \rightarrow 1$ as a limit. At the same time, if we
keep the continued spacetime dimension $D$ generic, the integral is
ultraviolet convergent for $\gamma >D/4$.

The Euclidean result 
\begin{equation}
B_{1\text{E}}(p_{\text{E}}^{2})=c(p_{\text{E}}^{2})^{D/2-2\gamma },\qquad c=%
\frac{\lambda ^{2}\Gamma \left( 2\gamma -\frac{D}{2}\right) \Gamma
^{2}\left( \frac{D}{2}-\gamma \right) }{2(4\pi )^{D/2}\Gamma ^{2}(\gamma
)\Gamma \left( D-2\gamma \right) },  \label{cc}
\end{equation}%
leads to the Minkowskian outcome\footnote{%
We recall that a factor $i$ must be included when the 1PI correlation
functions are switched from the Euclidean framework to the Minkowski one in
momentum space. The analytic continuation is $B_{\text{M}}^{\text{an}%
}(p^{2})=iB_{\text{E}}(-p^{2}-i\epsilon )$. The average continuation gives (%
\ref{dd}).} 
\begin{eqnarray}
B_{1\text{M}}(p^{2}) &=&\frac{i}{2}\left[ B_{1\text{E}}(-p^{2}-i\epsilon
)+B_{1\text{E}}(-p^{2}+i\epsilon )\right]  \notag \\
&=&ic\left[ \theta (-p^{2})(-p^{2})^{D/2-2\gamma }+\cos \left( \frac{D\pi }{2%
}-2\pi \gamma \right) \theta (p^{2})(p^{2})^{D/2-2\gamma }\right] .
\label{dd}
\end{eqnarray}%
The resummed (dressed) propagator reads%
\begin{equation*}
G_{1}^{\text{d}}(p^{2})=-\frac{i\theta (-p^{2})}{(-p^{2})^{\gamma
}-c(-p^{2})^{D/2-2\gamma }}-\frac{i\theta (p^{2})\cos \left( \pi \gamma
\right) }{(p^{2})^{\gamma }-c\cos \left( \pi \gamma \right) \cos \left( 
\frac{D\pi }{2}-2\pi \gamma \right) (p^{2})^{D/2-2\gamma }}.
\end{equation*}%
We see that, consistent with unitarity in the fakeon approach, the result is
purely imaginary.\ No physical degrees of freedom are on the external legs
(by projection) and none is turned on by the radiative corrections (by the
fakeon diagrammatics).

In the case of $\mathcal{L}_{2}$ we must search for a different formulation.

\subsection{Decomposition approach}

The decomposition approach consists of expressing the fractional propagator
as a superposition of local propagators with complex poles, and treating
those as fakeons inside diagrams. It can be applied to both theories (\ref%
{two}). In the case of $\mathcal{L}_{1}$, it leads to the same results as
the direct fakeon approach.

For $0<\gamma <1$ we can express the Euclidean $\mathcal{L}_{1}$ propagator
by means of the spectral decomposition \cite{cal}%
\begin{equation}
G_{1\text{E}}(p_{\text{E}}^{2})=\frac{1}{(p_{\text{E}}^{2})^{\gamma }}%
=\int_{0}^{\infty }\mathrm{d}s\frac{\rho _{\gamma }(s)}{p_{\text{E}}^{2}+s}%
,\qquad \rho _{\gamma }(s)=\frac{\sin \left( \pi \gamma \right) }{\pi
s^{\gamma }}.  \label{p1}
\end{equation}%
This expression is divergent for $1<\gamma <2$, where we can use%
\begin{equation}
G_{1}(p_{\text{E}}^{2})=\int_{0}^{\infty }\mathrm{d}s\rho _{\gamma
}(s)\left( \frac{1}{p_{\text{E}}^{2}+s}-\frac{1}{p_{\text{E}}^{2}}\right)
\label{p2}
\end{equation}%
instead.

The diagrams must be evaluated in the Euclidean framework, keeping the $s$%
-parameters fixed. Then one performs the average continuation to Minkowski
spacetime on the integrands of the $s$-integrals. Finally, one integrates
over the $s$-parameters.

Alternatively, one can apply the fakeon diagrammatics of refs. \cite%
{diagrammarMio,PVP20} directly in Minkowski spacetime. This means:

1. Calculate the usual Feynman diagrams with the $i\epsilon $ propagators,
i.e., 
\begin{equation}
\int_{0}^{\infty }\mathrm{d}s\frac{i\rho _{\gamma }(s)}{p^{2}-s+i\epsilon }%
,\qquad \int_{0}^{\infty }\mathrm{d}s\rho _{\gamma }(s)\left( \frac{i}{%
p^{2}-s+i\epsilon }-\frac{i}{p^{2}+i\epsilon }\right) ,  \label{r1}
\end{equation}%
for the cases $0<\gamma <1$ and $1<\gamma <2$, respectively, and combine
them in a prescribed way (which can be found in \cite{diagrammarMio,PVP20})
with additional diagrams built with (\ref{r1}) and the \textquotedblleft cut
propagators\textquotedblright 
\begin{equation}
\int_{0}^{\infty }\mathrm{d}s\rho _{\gamma }(s)(2\pi )\theta (\pm
p^{0})\delta (p^{2}-s),\qquad \int_{0}^{\infty }\mathrm{d}s\rho _{\gamma
}(s)(2\pi )\theta (\pm p^{0})\left[ \delta (p^{2}-s)-\delta (p^{2})\right] .
\label{r2}
\end{equation}

2. Evaluate the diagrams at fixed parameters $s$, using the dimensional
regularization and keeping the continued dimension $D$ generic; this means
that no renormalization and no expansion of $D$ around the physical
dimension $d$ must be performed at this stage (see below for examples in
calculations).

3. Evaluate the integrals on the $s$-parameters.

4. Expand $D$ around the physical value $d$ and, if needed, renormalize the
divergent parts.

In particular, the free two-point functions (\ref{G1}) are%
\begin{equation}
G_{1}(p^{2})=\mathcal{P}\int_{0}^{\infty }\mathrm{d}s\rho _{\gamma }(s)\frac{%
i}{p^{2}-s},\qquad G_{1}(p^{2})=\mathcal{P}\int_{0}^{\infty }\mathrm{d}s\rho
_{\gamma }(s)\left( \frac{i}{p^{2}-s}-\frac{i}{p^{2}}\right) ,
\label{freeModel1}
\end{equation}%
for $0<\gamma <1$ and $1<\gamma <2$, respectively.

Note that the sign of the spectral density is not relevant to the discussion
of this paper, since fakeons can accommodate both signs.

It is important that the subtraction of counterterms (if needed) not be
performed at fixed parameters $s$, since the $s$-integrals may generate
additional divergences or cancel out the spurious ones. Yet, the finite
parts of the diagrams may turn out to be correct nonetheless. In the next
section we illustrate these points explicitly in the bubble case.

A fact to keep in mind is that the fakeon diagrammatics does not amount to
computing Feynman diagrams with an alternative propagator (which would be (%
\ref{p1}) or (\ref{p2}) in the decomposition approach, (\ref{G1}) in the
direct approach). This option has been studied in the past \cite{Rocca} and
leads to violations of unitarity. Details on the difference between the two
approaches in ordinary, non-fractional models (which is already visible in
the bubble diagram) can be found in ref. \cite{Wheelerons}.

For $\mathcal{L}_{2}$ we write 
\begin{equation}
G_{2\text{E}}(p^{2})=\frac{1}{\left[ (p_{\text{E}}^{2})^{2}\right] ^{\gamma
/2}}=\int_{0}^{\infty }\mathrm{d}s\frac{\rho _{\gamma /2}(s)}{(p_{\text{E}%
}^{2})^{2}+s}.  \label{G2}
\end{equation}%
Note that $0<\gamma /2<1$ implies that the integral is IR\ and UV
convergent. The calculation of diagrams proceeds as described above.

In general, a fakeon decomposition, such as (\ref{p1})-(\ref{p2}) and (\ref%
{G2}), and the average continuation are operations that may not commute.
This is already evident at tree level, since $\mathcal{L}_{1}$ and $\mathcal{%
L}_{2}$ coincide in Euclidean space (the functions $-i/z^{\gamma }$ and $%
-i/(z^{2})^{\gamma /2}$ being identical for Re$[z]>0$), but lead to
different propagators in Minkowski spacetime, i.e., (\ref{G1}) versus%
\begin{equation}
G_{2}(p^{2})=-\frac{i}{\left[ (-p^{2})^{2}\right] ^{\gamma /2}}.  \label{g2}
\end{equation}%
Thus, caution must be used before commuting the two operations at the level
of diagrams. In the next section we prove that $\mathcal{L}_{1}$ and $%
\mathcal{L}_{2}$ give different results for the bubble diagram.

\subsubsection*{Model 1 and Model 2 face to face}

As said, the two models are identical in Euclidean space, yet different in
Minkowski spacetime. This discrepancy arises from the \textit{continuous}
sector of the decomposition, which can be dealt with in an infinite number
of ways. The outcome is an infinite family of inequivalent theories.

Let us summarize the situation so far, focusing on the case $0<\gamma <1$
for simplicity. In Euclidean space we have%
\begin{equation}
\hspace{0.01in}\frac{1}{(p_{\text{E}}^{2})^{\gamma }}=\int_{0}^{\infty }%
\mathrm{d}s\frac{\rho _{\gamma }(s)}{p_{\text{E}}^{2}+s},\qquad \frac{1}{%
\left[ (p_{\text{E}}^{2})^{2}\right] ^{\gamma /2}}=\int_{0}^{\infty }\mathrm{%
d}s\frac{\rho _{\gamma /2}(s)}{(p_{\text{E}}^{2})^{2}+s}.  \label{deco2}
\end{equation}%
The two expressions coincide for $z=p_{\text{E}}^{2}>0$, but give different
results when we turn to Minkowski spacetime (by performing the substitution $%
p_{\text{E}}^{2}\rightarrow -p^{2}-i\epsilon $ plus the average
continuation). If we choose the density $\rho _{\gamma }(s)$, we have a
continuum of fakeons with square masses equal to $s$. If we choose the
density $\rho _{\gamma /2}(s)$, we have a continuum of complex conjugate
fakeon pairs with square masses equal to $\pm i\sqrt{s}$.

Now, consider the formula 
\begin{equation}
\frac{1}{[(p_{\text{E}}^{2})^{n}]^{\gamma /n}}=\int_{0}^{\infty }\mathrm{d}s%
\frac{\rho _{\gamma /n}(s)}{(p_{\text{E}}^{2})^{n}+s},\qquad n\in \mathbb{N}%
_{+}.  \label{ndeco}
\end{equation}%
The left-hand sides of this equality are all equal to $(p_{\text{E}%
}^{2})^{-\gamma }$ in Euclidean space. Yet, different values of $n$ lead to
different Minkowskian theories.

For example, the Minkowski expressions%
\begin{equation}
\int_{0}^{\infty }\mathrm{d}s\frac{\rho _{\gamma /n}(s)}{(-p^{2}-i\epsilon
)^{n}+s}=\frac{1}{[(-p^{2}-i\epsilon )^{2k+1}]^{\gamma /(2k+1)}}
\label{decon}
\end{equation}%
differ from one another for odd $n=2k+1$, $k\in \mathbb{N}$, to the extent
that the average continuation%
\begin{equation}
\left. \frac{1}{[(-p^{2}-i\epsilon )^{2k+1}]^{\gamma /(2k+1)}}\right\vert _{%
\text{f}}=\frac{\theta (-p^{2})}{(-p^{2})^{\gamma }}+\cos \left( \frac{\pi
\gamma }{2k+1}\right) \frac{\theta (p^{2})}{(p^{2})^{\gamma }}
\label{ndecop}
\end{equation}%
(where the subscript \textquotedblleft f\textquotedblright\ means
\textquotedblleft fakeon\textquotedblright ) is $k$ dependent.

Note that in the complex plane $p_{\text{E}}^{2}\rightarrow z=\rho \mathrm{e}%
^{i\theta }$ the function on the left-hand side of (\ref{ndeco}) has cuts in
the radii $\theta =$ odd multiple of $\pi /n$. This means that the plane is
divided into disjoint regions for all $n>1$, which makes the direct fakeon
approach applicable only for $n=1$.

For all even values of $n$, formulas (\ref{decon}) give the same result,
which is (\ref{G2}) times $i$. Nevertheless, the calculations of the next
section (where we show that the options (\ref{deco2}) give different results
for the bubble diagram) suggest that the radiative corrections are all
different in those cases as well.

The conclusion is that different decompositions lead to different theories.
Since infinitely many options are available, infinitely many inequivalent
fractional theories correspond to the same Euclidean model.

\subsection{Non-Hermitian formulation}

The most natural option to make sense of kinetic terms involving fractional
powers of the D'Alembertian operator is to decompose the propagator as a
continuum of standard particles, rather than fakeons, treating the poles by
means of the Feynman $i\epsilon $ prescription \cite{CalcaRach}. We have
left this approach for last because it does not satisfy our requirements.

Specifically, the strategy amounts to using the K\"{a}ll\'{e}n--Lehmann
representation 
\begin{equation*}
G_{3}(p^{2})=-\frac{i}{(-p^{2}-i\epsilon )^{\gamma }}=i\int_{0}^{\infty }%
\mathrm{d}s\frac{\rho _{\gamma }(s)}{p^{2}-s+i\epsilon }
\end{equation*}%
(still at $0<\gamma <1$). The resulting theory is unitary, since the
Cutkosky-Veltman equations are obeyed. However, the classical equations of
motion%
\begin{equation*}
(\Box -i\epsilon )^{\gamma }\phi +\frac{\lambda }{2}\phi ^{2}=0
\end{equation*}%
are not real. Because of this, we do not pursue this formulation further in
this paper.

\section{Diagrammatics}

\label{diagra}\setcounter{equation}{0}

In this section we study the diagrammatics of fractional models and show
that different formulations lead to different results for the same
contribution to an amplitude.

As recalled, the fakeon diagrammatics involves a combination of Feynman
diagrams and additional diagrams built with both Feynman and cut
propagators. In the case of the bubble, the procedure amounts to applying a
simple operation (\textquotedblleft fakeon prescription\textquotedblright )
on the usual integral. Precisely, one starts from the Euclidean loop
integral and \textquotedblleft average-continues\textquotedblright\ it to
Minkowski spacetime, i.e., averages the analytic continuations around the
branch cut.

In the direct fakeon approach, one can apply the average continuation on the
Euclidean result (\ref{cc}), which gives (\ref{dd}). In the decomposition
approach, one has to apply it for fixed parameters $s$ and integrate over
them at the end. This is what we do explicitly in the next subsections.

\subsection{Model 1}

Let us take $0<\gamma <1$ for the moment. Using the left decomposition of (%
\ref{r1}) the bubble diagram gives the loop integral 
\begin{equation}
\frac{\lambda ^{2}}{2}\int_{0}^{\infty }\mathrm{d}s\hspace{0.01in}\rho
_{\gamma }(s)\int_{0}^{\infty }\mathrm{d}\sigma \hspace{0.01in}\rho _{\gamma
}(\sigma )\int \frac{\mathrm{d}^{D}k}{(2\pi )^{D}}\frac{1}{%
(k^{2}-s+i\epsilon )((p+k)^{2}-\sigma +i\epsilon )}.  \label{last}
\end{equation}%
The fakeon diagrammatics at fixed $s$ and $\sigma $ demands to average this
expression with the one obtained by replacing $+i\epsilon $ with $-i\epsilon 
$. We discuss this later.

Using Feynman parameters, formula (\ref{last}) gives%
\begin{equation}
\frac{i\lambda ^{2}}{2}\frac{\Gamma \left( 2-\frac{D}{2}\right) }{(4\pi
)^{D/2}}\int_{0}^{1}\mathrm{d}x\int_{0}^{\infty }\mathrm{d}s\hspace{0.01in}%
\rho _{\gamma }(s)\int_{0}^{\infty }\mathrm{d}\sigma \hspace{0.01in}\rho
_{\gamma }(\sigma )(-p^{2}x(1-x)+sx+\sigma (1-x)-i\epsilon )^{D/2-2}.
\label{bub}
\end{equation}%
The Schwinger representation%
\begin{equation*}
\frac{1}{A^{\alpha }}=\frac{1}{\Gamma (\alpha )}\int_{0}^{\infty }\mathrm{d}t%
\hspace{0.01in}t^{\alpha -1}\mathrm{e}^{-tA},\qquad \text{Re}[A]>0\text{, Re}%
[\alpha ]>0,
\end{equation*}%
allows us to write it as%
\begin{equation*}
\frac{i\lambda ^{2}(-i)^{D/2-2}}{2(4\pi )^{D/2}}\int_{0}^{1}\mathrm{d}%
x\int_{0}^{\infty }\mathrm{d}t\hspace{0.01in}t^{1-\frac{D}{2}}\mathrm{e}%
^{-t\epsilon +ip^{2}tx(1-x)}\int_{0}^{\infty }\mathrm{d}s\hspace{0.01in}\rho
_{\gamma }(s)\mathrm{e}^{-itsx}\int_{0}^{\infty }\mathrm{d}\sigma \hspace{%
0.01in}\rho _{\gamma }(\sigma )\mathrm{e}^{-it\sigma (1-x)},
\end{equation*}%
after which the $s$- and $\sigma $-integrals are straightforward. Using the
identity%
\begin{equation}
\int_{0}^{\infty }\mathrm{d}s\hspace{0.01in}\rho _{\gamma }(s)\mathrm{e}%
^{-itsx}=\frac{(itx)^{\gamma -1}}{\Gamma (\gamma )},\qquad 0<\gamma <1,
\label{sint}
\end{equation}%
we obtain 
\begin{equation*}
\frac{i\lambda ^{2}(-i)^{D/2}i^{2\gamma }}{2(4\pi )^{D/2}\Gamma ^{2}(\gamma )%
}\int_{0}^{1}\mathrm{d}x\hspace{0.01in}x^{\gamma -1}(1-x)^{\gamma
-1}\int_{0}^{\infty }\mathrm{d}t\hspace{0.01in}t^{2\gamma -1-\frac{D}{2}}%
\mathrm{e}^{-t\epsilon +itp^{2}x(1-x)}.
\end{equation*}%
At this point, the $t$-integral (defined by taking advantage of the
dimensional regularization in $D$) yields%
\begin{equation*}
\frac{i\lambda ^{2}}{2(4\pi )^{D/2}}\frac{\Gamma \left( 2\gamma -\frac{D}{2}%
\right) }{\Gamma ^{2}(\gamma )}(-p^{2}-i\epsilon )^{\frac{D}{2}-2\gamma
}\int_{0}^{1}\mathrm{d}x\hspace{0.01in}(x(1-x))^{\frac{D}{2}-\gamma -1}
\end{equation*}%
having rescaled $\epsilon $ by $x(1-x)$. Finally, the $x$ integral gives%
\begin{equation*}
\frac{i\lambda ^{2}}{2(4\pi )^{D/2}}\frac{\Gamma \left( 2\gamma -\frac{D}{2}%
\right) \Gamma ^{2}\left( \frac{D}{2}-\gamma \right) }{\Gamma ^{2}(\gamma
)\Gamma \left( D-2\gamma \right) }(-p^{2}-i\epsilon )^{\frac{D}{2}-2\gamma }.
\end{equation*}%
Once we sum the expression with $-p^{2}-i\epsilon \rightarrow
-p^{2}+i\epsilon $ and divide by 2, we find agreement with (\ref{cc}).

For $1<\gamma <2$, we use the right decomposition of (\ref{r1}) and proceed
in the same way. Instead of (\ref{sint}), we need the identity%
\begin{equation}
\int_{0}^{\infty }\mathrm{d}s\hspace{0.01in}\rho _{\gamma }(s)\left( \mathrm{%
e}^{-itsx}-1\right) =\frac{(itx)^{\gamma -1}}{\Gamma (\gamma )},\qquad
1<\gamma <2,  \label{spint}
\end{equation}%
and the final result is the same.

The case $\gamma \rightarrow 1$ can be reached as a limit. There, we have to
subtract the usual ultraviolet divergence in four dimensions.

Before proceeding, we highlight some potential pitfalls. In the
decomposition approach the bubble diagram of the fractional model is viewed
as a continuum of ordinary bubble diagrams. However, the latter are
divergent in $D=4$ (check the last integral of (\ref{last})), while the
former is convergent for every $\gamma \neq 1$. It is risky to expand $%
D=4-\varepsilon $ around its physical value $d=4$ and throw away the
divergent part in $\varepsilon $ \textit{before} integrating over $s$ and $%
\sigma $. Those integrals may generate divergent parts that are out of
control. Yet, if they are independent of the external momenta, as occurs in
the case of the bubble diagram, the finite part turns out to be right.

Let us verify this explicitly. We expand the integrand of (\ref{bub}) for $%
\varepsilon $ small, throw away the pole $1/\varepsilon $ and focus on the
convergent part, as we would normally do. Then we get%
\begin{equation}
-\frac{i\lambda ^{2}}{2(4\pi )^{2}}\int_{0}^{1}\mathrm{d}x\int_{0}^{\infty }%
\mathrm{d}s\hspace{0.01in}\rho _{\gamma }(s)\int_{0}^{\infty }\mathrm{d}%
\sigma \hspace{0.01in}\rho _{\gamma }(\sigma )\ln (-p^{2}x(1-x)+sx+\sigma
(1-x)-i\epsilon ).  \label{pint}
\end{equation}%
The $s$- and $\sigma $-integrations diverge, but their divergent parts are $%
p $-independent. We get rid of them by differentiating (\ref{pint}) with
respect to $p^{2}$, using the procedure outlined above (Schwinger
representation plus identities (\ref{sint}) and (\ref{spint})), integrating
back over $p^{2}$ and ignoring the $p$-independent additive constants. We
obtain%
\begin{equation}
\frac{i\lambda ^{2}}{2(4\pi )^{2}}\frac{\Gamma ^{2}(2-\gamma )\Gamma
(2\gamma -2)}{\Gamma ^{2}(\gamma )\Gamma (2\gamma )}(-p^{2}-i\epsilon
)^{2-2\gamma },  \label{loga}
\end{equation}%
which matches (\ref{cc}) for $D=4$.

Since we know that there are no divergent parts for $\gamma \neq 1$, and the
result must be proportional to $(-p^{2}-i\epsilon )^{2-2\gamma }$ in $D=4$
(on dimensional grounds), we conclude that (\ref{loga}) is complete: the
additive constant we are missing is equal to zero.

What has happened? The corrections of higher orders in $\varepsilon $
(originated by the expansion of $D$ around $d=4$) have the form (\ref{pint})
with powers of $\varepsilon $ in front and higher powers of the logarithm
inside. All such integrals are convergent as soon as they are differentiated
with respect to $p^{2}$. Hence, their contributions are killed by the limit $%
\varepsilon \rightarrow 0$.

Ultimately, the sector we are missing with this procedure is made of $p$%
-independent divergences, so the $p$-dependence of the finite part turns out
to be correct. Moreover, the divergent parts must sum to zero, since the
original integral was convergent.

Note that curious interplay between analytic \cite{anareg} and dimensional 
\cite{dimreg} regularization techniques. The former views $\gamma $ as a
complex exponent. A loop integral is evaluated in an open set of the $\gamma 
$ plane where it is convergent. Then the result is analytically continued to
the physical value $\gamma ^{\ast }$ of $\gamma $. The divergent parts are
negative powers of $\gamma -\gamma ^{\ast }$.

In summary, the Lagrangian $\mathcal{L}_{1}$ admits two equivalent
formulations, which also provide two options for the calculations: the
direct fakeon approach and the decomposition approach. A drawback of the
latter is that some caution must be used in order to properly deal with the
(possibly spurious) divergent parts.

\subsection{Model 2}

In Model 2 we cannot apply the direct fakeon approach, so we rely on the
decomposition strategy. The method of calculation described above extends
straightforwardly, but calculations are more involved.

Let us highlight the main question we need to answer. In formulas (\ref{cc})
and (\ref{dd}) we had a single cut on the complex plane, so the average
continuation was straightforward. Here, instead, we have a single cut on the
complex plane at fixed $s$ and $\sigma $, but the integrals with respect to
those variables can generate further cuts that divide the complex $p^{2}$
plane into disjoint regions. Hence, instead of (\ref{dd}), we expect a
Minkowskian result of the form 
\begin{equation}
B_{2\text{M}}(p^{2})=ic\left[ \theta (-p^{2})(-p^{2})^{D/2-2\gamma
}+g(\gamma )\theta (p^{2})(p^{2})^{D/2-2\gamma }\right] ,  \label{ee}
\end{equation}%
where $g(\gamma )$ is a new function of $\gamma $. We may even have a
different coefficient $g(\gamma )$ for every choice of $n$ in (\ref{ndeco}).
The results we derive below show precisely this. In particular, the case $n=2
$ does not give the $n=1$ result (\ref{dd}), which is $g(\gamma )=\cos
\left( \pi D/2-2\pi \gamma \right) $.

The $n=2$ Euclidean loop integral is 
\begin{equation}
B_{2\text{E}}(p_{\text{E}}^{2})=\frac{\lambda ^{2}}{2}\int_{0}^{\infty }%
\mathrm{d}s\hspace{0.01in}\rho _{\gamma /2}(s)\int_{0}^{\infty }\mathrm{d}%
\sigma \hspace{0.01in}\rho _{\gamma /2}(\sigma )\int \frac{\mathrm{d}^{D}k_{%
\text{E}}}{(2\pi )^{D}}\frac{1}{(k_{\text{E}}^{2})^{2}+s}\frac{1}{((p_{\text{%
E}}+k_{\text{E}})^{2})^{2}+\sigma }.  \label{BpE}
\end{equation}%
Here there are no ultraviolet divergences at fixed $s$ and $\sigma $ in $D<8$%
, so we can work directly in the physical dimension $d$.

Clearly, the Euclidean result is still (\ref{cc}). Hence for $p^{2}<0$ the
Minkowski one is%
\begin{equation}
B_{2\text{M}}(p^{2})=iB_{2\text{E}}(-p^{2})=iB_{1\text{E}}(-p^{2})\qquad
(p^{2}<0)\text{.}  \label{ccc}
\end{equation}

After replacing $s$ and $\sigma $ with $s^{2}$ and $\sigma ^{2}$ in (\ref%
{BpE}), respectively, we find%
\begin{equation}
B_{2\text{E}}(p_{\text{E}}^{2})=\frac{\lambda ^{2}}{2(4\pi )^{2}}%
\int_{0}^{\infty }\!\!\!\!\mathrm{d}s\hspace{0.01in}\rho _{\gamma
/2}(s^{2})\int_{0}^{\infty }\!\!\!\!\mathrm{d}\sigma \hspace{0.01in}\rho
_{\gamma /2}(\sigma ^{2})\int_{0}^{1}\!\!\mathrm{d}x\hspace{0.01in}\ln \frac{%
(p_{\text{E}}^{2})^{2}x^{2}(1-x)^{2}+(xs+(1-x)\sigma )^{2}}{(p_{\text{E}%
}^{2})^{2}x^{2}(1-x)^{2}+(xs-(1-x)\sigma )^{2}}  \label{bb}
\end{equation}%
in $d=4$ and 
\begin{eqnarray}
B_{2\text{E}}(p_{\text{E}}^{2}) &=&-\frac{\lambda ^{2}}{8\pi }%
\int_{0}^{\infty }\mathrm{d}s\hspace{0.01in}\rho _{\gamma
/2}(s^{2})\int_{0}^{\infty }\mathrm{d}\sigma \hspace{0.01in}\rho _{\gamma
/2}(\sigma ^{2})\int_{0}^{1}\mathrm{d}x\hspace{0.01in}\left[ \frac{1}{p_{%
\text{E}}^{2}x(1-x)-ixs-i(1-x)\sigma }\right.  \notag \\
&&\qquad \qquad \qquad \qquad \qquad \qquad \left. -\frac{1}{p_{\text{E}%
}^{2}x(1-x)-ixs+i(1-x)\sigma }+\text{c.c.}\right]  \label{bb2}
\end{eqnarray}%
in $d=2$.

To simplify the calculation enough without missing the main point, we study
the case $d=2$ for $1/2<\gamma <1$, where the integral of each contribution
in the brackets of (\ref{bb2}) is separately convergent. An analogous
procedure can be applied to the case $d=4$ after differentiating (\ref{bb})
with respect to $p_{\text{E}}^{2}$. There, however, the integrals of the
various contributions do not converge separately.

The next step is to make the average continuation of (\ref{bb2}) at fixed $s$
and $\sigma $. We give two procedures to achieve this goal: the first one,
done below, amounts to studying the $x$-integral without calculating it. The
second one, done in appendix \ref{B}, amounts to continuing the result of
the $x$-integral.

For $s$ and $\sigma $ generic, the expression%
\begin{equation}
f(z,s,\sigma )=\int_{0}^{1}\mathrm{d}x\hspace{0.01in}\left[ \frac{1}{%
zx(1-x)-ixs+i(1-x)\sigma }+\frac{1}{zx(1-x)+ixs-i(1-x)\sigma }\right]
\label{inti}
\end{equation}%
defines an analytic function of $z$ in the half plane Re$[z]>0$. Note that $%
f(z,s,\sigma )$ is real for real $z$, and symmetric with respect to the
exchange $s\leftrightarrow \sigma $. Finally, it is odd in $z$:%
\begin{equation}
f(-z,s,\sigma )=-f(z,s,\sigma ).  \label{sim}
\end{equation}

We consider the contributions in the brackets separately. Each one is
regular for Re[$z]\neq 0$. Singularities may occur when $z$ is purely
imaginary.

It is convenient to divide the complex plane into the regions Re$[z]>0$ and
Re$[z]<0$. Starting from the half plane Re$[z]>0$ (which is the Euclidean
region), our goal is to analytically continue $f(z,s,\sigma )$ to the half
plane Re$[z]<0$. For definiteness, we switch from Re$[z]>0$ to Re$[z]<0$ on
the upper half plane.

The first denominator vanishes at the points 
\begin{equation*}
x_{\pm }=\frac{1}{2}-i\frac{s+\sigma }{2z}\pm \frac{i}{2z}\sqrt{(s+\sigma
+iz)^{2}-4iz\sigma }.
\end{equation*}%
When $z$ crosses the positive imaginary axis, only $x_{+}$ crosses the
integration domain $0\leqslant x\leqslant 1$, and it does so from below.
This means that the analytic continuation of $f(z,s,\sigma )$ to Re$[z]<0$
is $f(z,s,\sigma )$ plus the contribution of the residue at $x_{+}$.

Something similar happens with the second denominator. At the end, we have
the analytic continuation 
\begin{equation}
f^{\text{an}}(z,s,\sigma )=\left\{ 
\begin{tabular}{ll}
$f(z,s,\sigma )$\qquad & for Re$[z]>0$ (Euclidean result)$,$ \\ 
$-f(-z,s,\sigma )+h(z,s,\sigma )$\qquad & for Re$[z]<0,$%
\end{tabular}%
\right.  \label{addi}
\end{equation}%
where%
\begin{equation}
h(z,s,\sigma )=\frac{2\pi }{\sqrt{(s+\sigma +iz)^{2}-4iz\sigma }}+\frac{2\pi 
}{\sqrt{(s+\sigma -iz)^{2}+4iz\sigma }}.  \label{hz}
\end{equation}%
The imaginary axis can be reached from both directions.

To calculate (\ref{bb2}), we also need to consider the integral 
\begin{equation}
g(z,s,\sigma )=\int_{0}^{1}\mathrm{d}x\hspace{0.01in}\left[ \frac{1}{%
zx(1-x)-ixs-i(1-x)\sigma }+\frac{1}{zx(1-x)+ixs+i(1-x)\sigma }\right]
\label{g}
\end{equation}%
for generic $s$ and $\sigma $. Again, $g(-z,s,\sigma )=-g(z,s,\sigma )$.

Studying the values of $z$ for which the denominators vanish in the $x$%
-integration domain, it is easy to check that $g(z,s,\sigma )$ is an
analytic function of $z$ in a \textquotedblleft safe\textquotedblright\ open
strip containing the real axis, extending by an amount equal to $(\sqrt{s}+%
\sqrt{\sigma })^{2}$ above and below the real axis itself. Hence, the
analytic continuation from the positive real axis to the negative one is
straightforward. We have 
\begin{equation}
g^{\text{an}}(z,s,\sigma )=\left\{ 
\begin{tabular}{ll}
$g(z,s,\sigma )$\qquad & for Re$[z]>0,$ \\ 
$-g(-z,s,\sigma )$ & for Re$[z]>0.$%
\end{tabular}%
\right.  \label{sim2}
\end{equation}

Finally, the analytic continuation of (\ref{bb2}) at fixed $s$ and $\sigma $
gives 
\begin{equation}
B_{2\text{E}}^{\text{an}}(z)=-\frac{\lambda ^{2}}{8\pi }\int_{0}^{\infty }%
\mathrm{d}s\hspace{0.01in}\rho _{\gamma /2}(s^{2})\int_{0}^{\infty }\mathrm{d%
}\sigma \hspace{0.01in}\rho _{\gamma /2}(\sigma ^{2})\left[ g^{\text{an}%
}(z,s,\sigma )-f^{\text{an}}(z,s,\sigma )\right] .  \label{b2ez}
\end{equation}%
Note that, although the integrand is analytic by construction, the integral $%
B_{2\text{E}}^{\text{an}}(z)$ need not be an analytic function.

An unexpected property, which emerges from the calculation, is that the
average continuation is automatically implemented by the integrals with
respect to $s$ and $\sigma $. In some sense, the decomposition approach
knows about fakeons by default. Due to this, the Minkowski bubble diagram
evaluates to%
\begin{equation}
B_{2\text{M}}(p^{2})=iB_{2\text{E}}^{\text{an}}(-p^{2})=\left\{ 
\begin{tabular}{ll}
$ic^{\prime }(-p^{2})\left[ (-p^{2})^{2}\right] ^{-\gamma }\qquad $ & for $%
p^{2}<0,$ \\ 
$-B_{2\text{M}}(-p^{2})+\Delta B_{2\text{M}}(p^{2})\quad $ & for $p^{2}>0,$%
\end{tabular}%
\right.  \label{b2m}
\end{equation}%
where%
\begin{equation}
c^{\prime }=\frac{\lambda ^{2}\Gamma \left( 2\gamma -1\right) \Gamma
^{2}\left( 1-\gamma \right) }{2(4\pi )\Gamma ^{2}(\gamma )\Gamma \left(
2-2\gamma \right) },\qquad \Delta B_{2\text{M}}(-z)=\frac{i\lambda ^{2}}{%
8\pi }\int_{0}^{\infty }\!\!\mathrm{d}s\hspace{0.01in}\rho _{\gamma
/2}(s^{2})\hspace{0.01in}\int_{0}^{\infty }\!\!\mathrm{d}\sigma \hspace{%
0.01in}\rho _{\gamma /2}(\sigma ^{2})\hspace{0.01in}h(z,s,\sigma ).
\label{disco}
\end{equation}%
On dimensional grounds, we know that the final result has the form (\ref{ee}%
).

We do not have the explicit result for $\Delta B_{2\text{M}}$, but for $%
p^{2}>0$ we can compare $B_{2\text{M}}(p^{2})$ numerically to (\ref{dd}) and
another simple option, which is 
\begin{equation*}
B_{2\text{M}}^{\prime }(p^{2})=ic^{\prime }\left[ \theta
(-p^{2})(-p^{2})^{1-2\gamma }+\theta (p^{2})(p^{2})^{1-2\gamma }\right]
=ic^{\prime }\left[ (p^{2})^{2}\right] ^{1/2-\gamma }.
\end{equation*}

\begin{figure}[t]
\begin{center}
\includegraphics[width=12truecm]{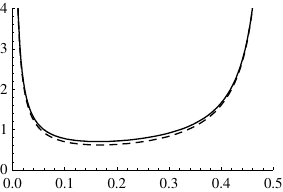}
\end{center}
\par
\vskip -.6truecm
\caption[Numerical comparison between $-2iB_{2\text{M}}(-1)/\protect\lambda %
^{2}$ (thick plot) and $-2iB_{2\text{M}}^{\prime }(-1)/\protect\lambda ^{2}$
(dashed plot) as functions of $\protect\delta =\protect\gamma -1/2$ for $0<%
\protect\delta <1/2$]{Numerical comparison between $-2iB_{2\text{M}}(1)/%
\protect\lambda ^{2}$ (thick plot) and $-2iB_{2\text{M}}^{\prime }(1)/%
\protect\lambda ^{2}$ (dashed plot) as functions of $\protect\delta =\protect%
\gamma -1/2$ for $0<\protect\delta <1/2$.}
\label{Contour}
\end{figure}

Fig. \ref{Contour} shows $-2iB_{2\text{M}}(p^{2})/\lambda ^{2}$ and $-2iB_{2%
\text{M}}^{\prime }(p^{2})/\lambda ^{2}$ for $p^{2}=1$ as functions of $%
\delta =\gamma -1/2$. The result is that the coefficient $g(\gamma )$ of
formula (\ref{ee}) is close to 1 for all values of $\gamma $ in the range we
are considering. However, it is not precisely 1. For example, we find $%
g(2/3)\simeq 1.45$, $g(3/4)\simeq 1.12$ and $g(5/6)\simeq 1.08$.

The difference with respect to (\ref{dd}) is more apparent, since $-\cos
\left( 2\pi \gamma \right) $ changes sign between $\gamma =1/2$ and $\gamma
=1$. 

Formulas (\ref{hz}), (\ref{b2m}) and (\ref{disco}) show that $\Delta B_{2%
\text{M}}$ is purely imaginary, which is what we expect from the average
continuation. Yet, we have derived it from $B_{2\text{E}}^{\text{an}}(z)$,
formula (\ref{b2ez}), which comes from the analytic continuation of $B_{2%
\text{E}}(z)$ at fixed $s$ and $\sigma $.

The fact, mentioned earlier, is that the analytic continuation of $B_{2\text{%
E}}(z)$ at fixed $s$ and $\sigma $ automatically turns into the average
continuation after integrating over $s$ and $\sigma $. Let us explain why.

By formula (\ref{hz}) the cuts of $h(z,s,\sigma )$ are located at $z=z_{\pm
} $, where%
\begin{equation*}
z_{\pm }=i(s-\sigma )\pm \sqrt{u+4s\sigma },\qquad u\geqslant 0.
\end{equation*}%
Only $z_{-}$ is relevant for us, since the function $h(z,s,\sigma )$
contributes just for Re[$z]<0$. This is a horizontal half line on the
complex $z$ plane, which crosses the real axis when $s-\sigma $ flips sign.
It is evident that the $s$- and $\sigma $-integrations average the
contributions where the squared momentum $p^{2}=-z\in \mathbb{R}$ is
symmetrically placed above and below the $z_{-}$ cut, which is what the
average continuation prescribes. Hence, (\ref{b2m}) is the final result.

Once again, we confirm that different decompositions lead to different
fractional quantum field theories, although the Euclidean counterpart is the
same.

Finally, combining (\ref{g2}) and (\ref{ee}), we find the dressed two-point
function 
\begin{equation*}
G_{2}^{\text{d}}(p^{2})=-\frac{i\theta (-p^{2})}{(-p^{2})^{\gamma
}-c(-p^{2})^{D/2-2\gamma }}-\frac{i\theta (p^{2})}{(p^{2})^{\gamma
}-cg(\gamma )(p^{2})^{D/2-2\gamma }}.
\end{equation*}

\subsubsection*{Remark}

We have found that the decomposition approach knows about fakeons by
default, in the form of the average continuation. At the computational
level, the reason is that the unusual, fractional propagator is replaced by
a continuum of standard, local propagators, thereby providing regions where
the cuts of the bubble diagram are placed both above and below the real
axis, prior to integrating over $s$ and $\sigma $. Hence, the integrations
average the analytic continuations from above and below, as per the average
continuation. Yet, it is remarkable that there is no need to enforce this as
an external prescription.

This raises several questions which deserve further scrutiny. For example,
it remains to be seen whether other diagrams exhibit the same property.
Another interesting direction is to investigate to what extent it makes
sense to view fakeons in local theories (where the prescription is a choice)
as limits of the \textquotedblleft natural fakeons\textquotedblright\ of
fractional theories when the fractional power approaches an integer.

\section{Extended models with a fractional d'Alembertian}

\label{extfrac}\setcounter{equation}{0}

So far we have focused on simple models like (\ref{two}), but it is clear
that the direct fakeon approach and the decomposition approach are general.
In this section, we illustrate extended models described by Lagrangians of
the forms%
\begin{equation*}
\mathcal{L}_{1}=-\frac{1}{2}\phi \Box \left( 1+\frac{\Box ^{\gamma }}{%
M^{2\gamma }}\right) \phi +\mathcal{L}_{\text{int}},\qquad \mathcal{L}_{2}=-%
\frac{1}{2}\phi \Box \left( 1+\left( \frac{\Box ^{2}}{M^{4}}\right) ^{\gamma
/2}\right) \phi +\mathcal{L}_{\text{int}},
\end{equation*}%
where $\gamma >0$, and $\mathcal{L}_{\text{int}}$ denotes generic self
interactions.

\subsection{Model 1}

As before, the classical $\mathcal{L}_{1}$ field equations%
\begin{equation}
\Box \left( 1+\frac{\Box ^{\gamma }}{M^{2\gamma }}\right) \phi =I_{\text{int}%
},  \label{boss}
\end{equation}%
where $I_{\text{int}}$ denotes interactions, are not acceptable, since they
are ill defined or not real. The box in front signals a physical degree of
freedom that we can keep as such. The rest of the kinetic operator on the
left-hand side must be treated by means of fakeons.

We write the Euclidean propagator as%
\begin{equation}
\frac{1}{p_{\text{E}}^{2}}-\frac{1}{M^{2}}F_{\gamma }(p_{\text{E}%
}^{2}/M^{2}),\qquad \text{where }F_{\gamma }(z)=\frac{z^{\gamma -1}}{%
1+z^{\gamma }}.  \label{Fg}
\end{equation}

In the direct fakeon approach we continue the Euclidean correlation
functions to Min\-kow\-ski spacetime as follows: all the thresholds, cuts,
poles, etc., that depend on the fractional sector encoded in $F_{\gamma }$
are treated by means of the average continuation, while those that are
independent of the fractional sector are treated by means of the $i\epsilon $
prescription. If $\gamma $ is a free parameter, the former can be
distinguished from the latter because they are $\gamma $ dependent.

In the case of the free propagator, we just need to apply the average
continuation to the function $F_{\gamma }$ and use the Feynman prescription
for the physical pole. We get%
\begin{eqnarray}
G_{1}(p^{2}) &=&\frac{i}{p^{2}+i\epsilon }+\frac{i}{M^{2}}\text{Re}\left[
F_{\gamma }(-p^{2}/M^{2}-i\epsilon )\right]  \notag \\
&=&\frac{i}{p^{2}+i\epsilon }+\frac{i\theta (-p^{2})(-p^{2})^{\gamma -1}}{%
M^{2\gamma }+(-p^{2})^{\gamma }}+\frac{i\cos (\pi \gamma )\theta
(p^{2})(p^{2})^{\gamma -1}}{M^{2\gamma }+(p^{2})^{\gamma }}.  \label{g1p}
\end{eqnarray}%
The field equations of the theory with fakeons are then%
\begin{equation*}
\Box \phi =\left. \frac{M^{2\gamma }}{M^{2\gamma }+\Box ^{\gamma }}%
\right\vert _{\text{f}}I_{\text{int}}=i\Box G_{1}(-\Box )I_{\text{int}}=%
\left[ 1-\frac{\theta (\Box )\Box ^{\gamma }}{M^{2\gamma }+\Box ^{\gamma }}+%
\frac{\cos (\pi \gamma )\theta (-\Box )(-\Box )^{\gamma }}{M^{2\gamma
}+(-\Box )^{\gamma }}\right] I_{\text{int}}.
\end{equation*}

The decomposition approach is applied as follows. For $0<\gamma <1$, the
function $F_{\gamma }(z)$ has no poles for $-\pi <$ Arg[$z$] $<\pi $ and a
cut along the negative real axis. Using the Cauchy theorem, we can integrate
along the cut and represent $F_{\gamma }(z)$ by means of the decomposition%
\begin{equation}
F_{\gamma }(z)=\int_{0}^{\infty }\frac{\rho _{\gamma }(s)\mathrm{d}s}{z+s}%
,\qquad \rho _{\gamma }(s)=\frac{1}{\pi }\frac{s^{\gamma -1}\sin (\pi \gamma
)}{1+s^{2\gamma }+2s^{\gamma }\cos (\pi \gamma )}.  \label{ros}
\end{equation}%
On the other hand, if 
\begin{equation*}
2k+1\leqslant \gamma <2k+3,\qquad k\in \mathbb{N},
\end{equation*}%
the function $F_{\gamma }(z)$ has poles at $z=\mathrm{e}^{\frac{i\pi }{%
\gamma }n}$, where $n$ is any odd integer between $-2k-1$ and $2k+1$. Then 
\begin{equation}
F_{\gamma }(z)=\int_{0}^{\infty }\mathrm{d}s\frac{\rho _{\gamma }(s)}{z+s}%
+R_{\gamma }(z),\qquad R_{\gamma }(z)\equiv \frac{1}{\gamma }\sum_{\substack{
n=-2k-1 \\ n=\text{odd}^{\ast }}}^{2k+1}\frac{1}{z-\mathrm{e}^{i\pi n/\gamma
}},  \label{deco}
\end{equation}%
where the star on \textquotedblleft odd\textquotedblright\ means that when $%
\gamma $ itself is odd ($\gamma =2k+1$), the contribution of $n=-2k-1$ is
ignored; otherwise, the pole $1/(z+1)$ is counted twice. It is easy to check
(\ref{deco}) in simple cases such as $\gamma \in \mathbb{N}_{+}$, $\gamma
=1/2$, $1/4$, $2/3...$, etc.

Diagrams are built as explained previously, using (\ref{deco}) and applying
the fakeon diagrammatics at fixed $s$ and $n$. The pole at $p^{2}=0$ is
treated by means of the $i\epsilon $ prescription, while the poles at $%
-z=p^{2}=M^{2}s$ and those at $-z=p^{2}=-M^{2}\mathrm{e}^{i\pi n/\gamma }$
are treated as fakeons. The sum over $n$ and the integral over $s$ are done
at the end. Note that there are no tachyons (for which the fakeon
prescription is not guaranteed to work).

In particular, the propagator (\ref{g1p}) is also equal to%
\begin{eqnarray}
G_{1}(p^{2}) &=&\frac{i}{p^{2}+i\epsilon }-\mathcal{P}\int_{0}^{\infty }%
\mathrm{d}s\frac{i\rho _{\gamma }(s)}{p^{2}-sM^{2}},  \notag \\
G_{1}(p^{2}) &=&\frac{i}{p^{2}+i\epsilon }-\mathcal{P}\int_{0}^{\infty }%
\mathrm{d}s\frac{i\rho _{\gamma }(s)}{p^{2}-sM^{2}}-\frac{1}{\gamma }%
\mathcal{P}\sum_{\substack{ n=-2k-1  \\ n=\text{odd}^{\ast }}}^{2k+1}\frac{i%
}{p^{2}+M^{2}\mathrm{e}^{i\pi n/\gamma }},  \label{g1p2}
\end{eqnarray}%
for $0<\gamma <1$ and $2k+1\leqslant \gamma <2k+3$, $k\in \mathbb{N}$,
respectively.

\subsection{Model 2}

In the $\mathcal{L}_{2}$ case, the classical field equations 
\begin{equation}
\Box \left( 1+\left( \frac{\Box ^{2}}{M^{4}}\right) ^{\gamma /2}\right) \phi
=I_{\text{int}}  \label{boss2}
\end{equation}%
are acceptable in the form they appear.

We write the Euclidean propagator as%
\begin{equation}
\frac{1}{p_{\text{E}}^{2}}\frac{1}{\left( 1+\frac{(p_{\text{E}}^{2})^{2}}{%
M^{4}}\right) ^{\gamma /2}}=\frac{1}{p_{\text{E}}^{2}}-\frac{p_{\text{E}}^{2}%
}{M^{4}}F_{\gamma /2}(u)=\frac{1}{p_{\text{E}}^{2}}\left[ 1-uF_{\gamma /2}(u)%
\right] ,  \label{odd}
\end{equation}%
where $u=(p_{\text{E}}^{2})^{2}/M^{4}$ and $F_{\gamma /2}(u)$ can be read
from (\ref{Fg}).

Again, we treat the pole at $p_{\text{E}}^{2}=-p^{2}=0$ as physical by means
of the Feynman $i\epsilon $ prescription and the remnant by means of the
fakeon approach at fixed $s$ and $n$, using the decomposition (\ref{deco}).

For $0<\gamma <2$ the Minkowskian propagator is%
\begin{equation}
G_{2}(p^{2})=\frac{i}{p^{2}+i\epsilon }-\int_{0}^{\infty }\mathrm{d}s\frac{%
ip^{2}\rho _{\gamma /2}(s)}{(p^{2})^{2}+sM^{4}},  \label{propp}
\end{equation}%
while for $4k+2\leqslant \gamma <4k+6$, $k\in \mathbb{N}$, it is%
\begin{equation}
G_{2}(p^{2})=\frac{i}{p^{2}+i\epsilon }-\int_{0}^{\infty }\mathrm{d}s\frac{%
ip^{2}\rho _{\gamma /2}(s)}{(p^{2})^{2}+sM^{4}}-\frac{2}{\gamma }\sum 
_{\substack{ n=-2k-1  \\ n=\text{odd}^{\ast }}}^{2k+1}\frac{ip^{2}}{%
(p^{2})^{2}-M^{4}\mathrm{e}^{2i\pi n/\gamma }},  \label{propa}
\end{equation}%
where, as before, $n$ is any odd integer number between $-2k-1$ and $2k+1$.
There is a continuum of fakeon poles at $p^{2}=\pm iM^{2}\sqrt{s}$ and a
discrete set of fakeon poles at $p^{2}=\pm M^{2}\mathrm{e}^{i\pi n/\gamma }$%
. They all come in complex conjugate pairs.

A few examples may help. In the case $\gamma =1$ we have%
\begin{equation*}
F_{1/2}(u)=\frac{1}{\sqrt{u}+u}=\int_{0}^{\infty }\frac{\mathrm{d}s}{\pi 
\sqrt{s}(1+s)}\frac{1}{s+u},
\end{equation*}%
while $\gamma =1/2$ gives%
\begin{equation*}
F_{1/4}(u)=\frac{1}{u^{3/4}+u}=\int_{0}^{\infty }\frac{\mathrm{d}s}{\sqrt{2}%
\pi s^{3/4}(1+\sqrt{2}s^{1/4}+\sqrt{s})}\frac{1}{s+u}.
\end{equation*}%
A simple situation where the decomposition with discrete fakeons can be
checked straightforwardly is the case $\gamma =3$, which gives%
\begin{eqnarray*}
F_{3/2}(u) &=&\frac{\sqrt{u}}{1+u^{3/2}}=f_{3/2}(u)+R_{3/2}(u),\qquad
R_{3/2}(u)=\frac{2}{3}\frac{1+2u}{1+u+u^{2}}, \\
f_{3/2}(u) &=&\int_{0}^{\infty }\mathrm{d}s\frac{\rho _{3/2}(s)}{s+u}=-\frac{%
2+\sqrt{u}}{3(1+\sqrt{u})(1+\sqrt{u}+u)}.
\end{eqnarray*}

The expression (\ref{propa}) can be used inside loop diagrams. A loop
integral is first evaluated at fixed $s$ and $n$ in the Euclidean framework.
Then one performs the average continuation to Minkowski spacetime, still at
fixed $s$ and $n$. Finally, one integrates over $s$ and sums on $n$.

For every $s$ and $n$, the loop integrals have the usual structure, the
integrands being ratios of polynomials of the momenta. It is also
straightforward to use the diagrammatic rules of \cite{diagrammarMio,PVP20}.

More general models can be studied along the same lines.

\subsection{Other formulations}

As said, the options are infinitely many and lead to inequivalent theories.
An alternative decomposition for Model 2 is studied in ref. \cite{Calcarest2}%
. The main difference is that our choice preserves the oddity of (\ref{odd})
under $z=p_{\text{E}}^{2}\rightarrow -z$, while the arrangement of \cite%
{Calcarest2} breaks the propagator into the sum of the physical pole plus an
even function of $p^{2}$.

Moreover, in the spirit of section \ref{frac} we point out an infinite class
of decompositions aligned with our choice (\ref{odd}), which are%
\begin{eqnarray*}
F_{\gamma /2}(u) &=&\frac{(u^{n})^{(\gamma -2)/(2n)}}{1+(u^{n})^{\gamma
/(2n)}}=\int_{0}^{\infty }\mathrm{d}s\frac{\rho _{n}(s)}{u^{n}+s},\qquad u>0,
\\
\rho _{n}(s) &=&\frac{s^{(\gamma -2)/(2n)}}{\pi }\frac{s^{\gamma /(2n)}\sin
(\pi /n)-\sin \left( \pi (\gamma -2)/(2n)\right) }{1+s^{\gamma
/n}+2s^{\gamma /(2n)}\cos \left( \pi \gamma /(2n)\right) },
\end{eqnarray*}%
for $0<\gamma <2n$. For the other values of $\gamma $ we need to add
discrete sets of fakeons as before.

\section{Fractional covariant d'Alembertians}

\label{covfrac}\setcounter{equation}{0}

In this section we extend the formulations to fractional powers of covariant
d'Alembertians. We focus on gauge theories, since gravity can be treated
similarly.

Consider the Lagrangians%
\begin{eqnarray*}
\mathcal{L}_{1} &=&-\bar{\phi}D_{\mu }D^{\mu }\left( 1+\frac{(D_{\nu }D^{\nu
})^{\gamma }}{M^{2\gamma }}\right) \phi +\mathcal{L}_{\text{int}}, \\
\mathcal{L}_{2} &=&-\bar{\phi}D_{\mu }D^{\mu }\left( 1+\left( \frac{(D_{\nu
}D^{\nu })^{2}}{M^{4}}\right) ^{\gamma /2}\right) \phi +\mathcal{L}_{\text{%
int}},
\end{eqnarray*}%
where $D_{\mu }=\partial _{\mu }-ieA_{\mu }$ is the covariant derivative.

Below we derive the vertices and explain how to build the correlation
functions. We summarize here the rules: $a$) in the direct approach to Model
1 we switch $\mathcal{L}_{1}$ to the Euclidean framework, then switch the
correlation functions back to Minkowski spacetime by means of the average
continuation; $b$) in the decomposition approach (for both $\mathcal{L}_{1}$
and $\mathcal{L}_{2}$) we also switch the Lagrangian to the Euclidean
framework, write the correlation functions by means the appropriate density $%
\rho (s)$ (with the possible addition of discrete sets of fakeons, labeled
by some integer $n$), average-continue the Euclidean loop integrals to
Minkowski spacetime at fixed $s$ and $n$, and finally integrate over $s$ and
sum over $n$.

\subsection{Vertices}

The densities can also be used as efficient tools to derive the vertices. We
first work in Model 2 then in Model 1. For definiteness, we assume $0<\gamma
<2$ for Model 2 and $0<\gamma <1$ for Model 1. In the other cases we have to
include discrete sets of fakeons.

The scalar propagator (\ref{propp}) leads to the covariantized expression%
\begin{equation}
-\frac{i}{D_{\mu }D^{\mu }-i\epsilon }+iD_{\nu }D^{\nu }\int_{0}^{\infty }%
\frac{\rho _{\gamma /2}(s)\hspace{0.01in}\mathrm{d}s}{(D_{\mu }D^{\mu
})^{2}+sM^{4}},  \label{coo}
\end{equation}%
where $\rho _{\gamma /2}(s)$ is the one of (\ref{ros}), which can be
expanded in powers of the coupling as follows. Focusing on the integral in
the last term of (\ref{coo}) and writing 
\begin{equation*}
(D_{\mu }D^{\mu })^{2}\equiv \Box ^{2}+\Upsilon ,\qquad \Upsilon =V_{\mu \nu
\rho }\partial ^{\mu }\partial ^{\nu }\partial ^{\rho }+V_{\mu \nu }\partial
^{\mu }\partial ^{\nu }+V_{\mu }\partial ^{\mu }+V,
\end{equation*}%
for some functions $V_{\mu \nu \rho }$, $V_{\mu \nu }$, $V_{\mu }$ and $V$,
the geometric series gives%
\begin{equation}
\int_{0}^{\infty }\frac{\rho _{\gamma /2}(s)\hspace{0.01in}\mathrm{d}s}{\Box
^{2}+sM^{4}+\Upsilon }=\sum_{m=0}^{\infty }\int_{0}^{\infty }\frac{\rho
_{\gamma /2}(s)\hspace{0.01in}\mathrm{d}s}{\Box ^{2}+sM^{4}}\left( -\Upsilon 
\frac{1}{\Box ^{2}+sM^{4}}\right) ^{m}.  \label{aF}
\end{equation}%
At this point, we define%
\begin{equation}
F_{x_{1},\ldots ,x_{m+1}}\equiv \int_{0}^{\infty }\frac{(-1)^{m}\rho
_{\gamma /2}(s)\hspace{0.01in}\mathrm{d}s}{\prod%
\limits_{i=1}^{m+1}(x_{i}+sM^{4})},\qquad m\in \mathbb{N},  \label{F}
\end{equation}%
and switch to momentum space. The Fourier transform of (\ref{aF}) with
momentum $p$ is%
\begin{equation}
\int \mathrm{d}^{4}x\int_{0}^{\infty }\frac{\rho _{\gamma /2}(s)\mathrm{d}s}{%
\Box ^{2}+sM^{4}+\Upsilon }\mathrm{e}^{ip\cdot x}=\sum_{m=0}^{\infty }\int 
\tilde{\Upsilon}(k_{1})\cdots \tilde{\Upsilon}(k_{m})F_{((p-\hat{k}%
_{m})^{2})^{2},\cdots ((p-\hat{k}_{1})^{2})^{2},(p^{2})^{2}},  \label{ups}
\end{equation}%
where $\hat{k}_{j}=k_{1}+\cdots +k_{j}$, while $k_{i}$ are the incoming
momenta of the $\tilde{\Upsilon}$ insertions (Fourier transforms of $%
\Upsilon $). The $k_{i}$-integration measures are understood.

Formula (\ref{ups}) shows both a compact expression (to the right) and\ a
decomposed expression (to the left). Compact expressions of this type can be
used in the direct approach to Model 1.

In the case of Model 1, we write%
\begin{equation*}
D_{\mu }D^{\mu }\equiv \Box +\Xi ,\qquad \Xi =-2ieA_{\mu }\partial ^{\mu
}-ie(\partial ^{\mu }A_{\mu })-e^{2}A_{\mu }A^{\mu }.
\end{equation*}%
The geometric series gives%
\begin{equation}
\int \mathrm{d}^{4}x\int_{0}^{\infty }\frac{\rho _{\gamma }(s)\mathrm{d}s}{%
\Box +s+\Xi -i\epsilon }\mathrm{e}^{ip\cdot x}=\sum_{m=0}^{\infty }\int 
\tilde{\Xi}(k_{1})\cdots \tilde{\Xi}(k_{m})H_{-(p-\hat{k}_{m})^{2}-i\epsilon
,\cdots -(p-\hat{k}_{1})^{2}-i\epsilon ,-p^{2}-i\epsilon },  \label{ups2}
\end{equation}%
where $H$ is $F$ with $\rho _{\gamma /2}(s)\rightarrow \rho _{\gamma }(s)$
and $M\rightarrow 1$. Then we apply the average continuation at fixed $s$.

\subsection{Gauge invariance}

Now we illustrate how gauge invariance works at the level of correlations
functions, through the Ward-Takahashi-Slavnov-Taylor (WTST) identities \cite%
{WTST}.

To keep formulas simple, we consider fractional scalar QED models with
Lagrangian%
\begin{equation}
\mathcal{L}_{1}=-\frac{1}{4}F_{\mu \nu }F^{\mu \nu }-\bar{\varphi}(D^{\mu
}D_{\mu })^{\gamma }\varphi ,\qquad \mathcal{L}_{2}=-\frac{1}{4}F_{\mu \nu
}F^{\mu \nu }-\bar{\varphi}[(D^{\mu }D_{\mu })^{2}]^{\gamma /2}\varphi ,
\label{Ls}
\end{equation}%
where $\varphi $ is a charged scalar field. The theories (\ref{Ls}) are
invariant under the infinitesimal gauge transformation $\delta \varphi
=ie\Lambda \varphi $, $\delta A_{\mu }=\partial _{\mu }\Lambda $.

\subsection{WTST\ identities: direct approach}

When needed, the switch to Euclidean space is understood. We comment on the
gauge invariance of the average continuation to Minkowski spacetime at the
end.

We begin with the direct approach to Model 1, where we can apply compact
formulas. Using the density of (\ref{p1}), the first order of (\ref{ups2})
in $\tilde{\Xi}$ gives%
\begin{equation}
\tilde{\Xi}(k)H_{-(p-k)^{2}-i\epsilon ,-p^{2}-i\epsilon }=\tilde{\Xi}(k)%
\frac{[-(p-k)^{2}-i\epsilon ]^{-\gamma }-(-p^{2}-i\epsilon )^{-\gamma }}{%
-(p-k)^{2}+p^{2}}.  \label{1st}
\end{equation}%
Next, writing 
\begin{equation*}
D_{\mu }D^{\mu }=\Box +\Xi =-p^{2}+eA_{\mu }(k)(2p_{\mu }-k_{\mu })+\mathcal{%
O}(A^{2})
\end{equation*}%
and acting on $\mathrm{e}^{ip\cdot x}$, we find that the vertex with one
gauge field is%
\begin{equation*}
\tilde{\Xi}(k)H_{-(p-k)^{2}-i\epsilon ,-p^{2}-i\epsilon }=eA_{\mu }(k)U^{\mu
}
\end{equation*}%
with%
\begin{equation}
U^{\mu }=-\int_{0}^{\infty }\frac{(2p^{\mu }-k^{\mu })\rho _{\gamma }(s)%
\mathrm{d}s}{((p-k)^{2}-s+i\epsilon )(p^{2}-s+i\epsilon )}=(2p^{\mu }-k^{\mu
})\frac{[-(p-k)^{2}-i\epsilon ]^{-\gamma }-(-p^{2}-i\epsilon )^{-\gamma }}{%
-(p-k)^{2}+p^{2}}.  \label{Vmu}
\end{equation}

The generalization of the usual Ward identity is obtained by replacing $%
A^{\mu }(k)$ with $-ik^{\mu }$. Using the right expression of (\ref{Vmu}) we
find%
\begin{equation}
-ik^{\mu }U_{\mu }=-\frac{i}{[-(p-k)^{2}-i\epsilon ]^{\gamma }}+\frac{i}{%
(-p^{2}-i\epsilon )^{\gamma }}.  \label{ward}
\end{equation}

If we move to Euclidean space, this formula gives the basic Euclidean Ward
identity. Indeed, the right-hand side turns into the difference between two
Euclidean propagators with the expected momenta. This proves that the
Euclidean correlation functions are gauge-invariant.

Moreover, the analytic continuation is a gauge-invariant operation (on
gauge-invariant Euclidean correlation functions), because the WTST
identities, such as (\ref{ward}), hold at the integrand level. During the
continuation the identities continue to hold simply because they are
\textquotedblleft identities\textquotedblright , hence the same operations
act on their left-hand sides as on their right-hand sides. They also hold
during the average continuation, because it is an average of analytic
continuations.

For example, when we average the $i\epsilon $ contributions (\ref{ward})
with its $-i\epsilon $ counterpart, the right-hand side gives the difference
between two propagators of Model 1 with the correct momenta, while the
average of $U_{\mu }$ gives the vertex of Model 1.

\subsection{WTST\ identities: decomposition approach}

As said, the key property of the WTST\ identities is that they can be proved
at the level of integrands, without the need to calculate any diagrams. In
the decomposition approach, they hold at fixed $s$ and $n$, where they look
like the WTST identities of ordinary theories.

Specifically, using the mid expression of (\ref{Vmu}) we easily find%
\begin{eqnarray*}
-ik_{\mu }(k)U^{\mu } &=&-i\left[ (p-k)^{2}-p^{2}\right] \int_{0}^{\infty }%
\frac{\rho _{\gamma }(s)\hspace{0.01in}\mathrm{d}s}{((p-k)^{2}-s+i\epsilon
)(p^{2}-s+i\epsilon )} \\
&=&\int_{0}^{\infty }\frac{i\rho _{\gamma }(s)\hspace{0.01in}\mathrm{d}s}{%
(p-k)^{2}-s+i\epsilon }-\int_{0}^{\infty }\frac{i\rho _{\gamma }(s)\hspace{%
0.01in}\mathrm{d}s}{p^{2}-s+i\epsilon }.
\end{eqnarray*}%
The second expression is the action on the decomposed vertices, while the
right-hand side is the difference of two decomposed propagators.

At the level of integrands we just have the usual Ward identity%
\begin{equation*}
-ik_{\mu }\frac{i}{(p-k)^{2}-s+i\epsilon }(2p_{\mu }-k_{\mu })\frac{i}{%
p^{2}-s+i\epsilon }=\frac{i}{(p+k)^{2}-s+i\epsilon }-\frac{i}{%
p^{2}-s+i\epsilon }.
\end{equation*}

One proceeds in a similar way in the case of Model 2.

\section{Classicization and degrees of freedom}

\label{class}\setcounter{equation}{0}

In this section we study the number of degrees of freedom of fractional
theories. In the classical limit we have two sources of nonlocality: the
fractional power of derivative operators and fakeons. This raises the
concern that an infinite number of initial conditions may be needed to
uniquely determine the solution of the field equations. We prove that it is
not so. Actually, only the expected, physical degrees of freedom propagate.

In nonfractional theories the issue was studied in ref. \cite{Calcagni},
where it was shown that fakeons are immune from the problem just stated. In
fractional models, we must investigate the impact of the additional
nonlocality due to the fractional power.

We illustrate the procedure in simple solvable examples. We assume that $%
\gamma $ is truly fractional, i.e., $\gamma =m/n$ with $m,n\in \mathbb{N}$, $%
n\neq 0$, and concentrate on quantum mechanics ($D=1$).

Since we are working in the realm of perturbative quantum field theory, it
is sufficient to count the number of degrees of freedom when interactions
are switched off. Interactions do not affect the counting, since they are
turned on iteratively. We focus on models of the extended classes considered
in section \ref{extfrac}, where the kinetic operator is nonsingular when the
fractional power is turned off. The models of section \ref{frac} are
singular in this respect. It is not known at present how to properly count
the number of degrees of freedom in cases like (\ref{two}) relying on
standard QFT techniques.

Consider equation (\ref{boss2}) with a harmonic force:%
\begin{equation}
\left( 1+\left( \frac{\mathrm{d}^{4}/\mathrm{d}t^{4}}{M^{4}}\right) ^{\gamma
/2}\right) \ddot{x}=-\omega ^{2}x.  \label{ew}
\end{equation}%
The correct equation of motion of the theory with fakeons is obtained by
inverting the operator between parentheses by means of the fakeon
prescription (see, for example, \cite{Calcagni}). We find 
\begin{equation}
\ddot{x}(t)=-\omega ^{2}\int_{-\infty }^{+\infty }\mathrm{d}t^{\prime }G_{%
\text{f}}^{\gamma }(t-t^{\prime })x(t^{\prime }),\qquad G_{\text{f}}^{\gamma
}(t)=\int_{-\infty }^{+\infty }\hspace{0.01in}\frac{\mathrm{d}\varpi }{2\pi }%
\frac{M^{2\gamma }\mathrm{e}^{-i\varpi t}}{M^{2\gamma }+(\varpi
^{2})^{\gamma }},  \label{nleq}
\end{equation}%
where $G_{\text{f}}^{\gamma }(t)$ denotes the fakeon Green function. For the
uses below we need its asymptotic behavior for $0<\gamma <1/2$, which is%
\begin{equation}
G_{\text{f}}^{\gamma }(t)\simeq \frac{\sin (\pi \gamma )\Gamma (1+2\gamma )}{%
\pi M^{2\gamma }|t|^{1+2\gamma }},  \label{asy}
\end{equation}%
and%
\begin{equation}
G_{\text{f}}^{1}(t)=\frac{M}{2}\mathrm{e}^{-M|t|}.  \label{asy2}
\end{equation}

Switching to Fourier transforms, we search for solutions to (\ref{ew}) of
the form%
\begin{equation}
x(t)=\sum_{j}c_{j}\mathrm{e}^{i\lambda _{j}tM},  \label{ct}
\end{equation}%
where $c_{j}$ are arbitrary coefficients. The frequencies $\lambda _{j}$
solve the equation%
\begin{equation}
\left[ 1+(\lambda _{j}^{4})^{\gamma /2}\right] \lambda _{j}^{2}=\tilde{\omega%
}^{2},  \label{cut}
\end{equation}%
where $\tilde{\omega}=\omega /M$. Squaring and then raising to the power $m$%
, we obtain the polynomial equations 
\begin{equation}
\left( 1+v_{j}\right) ^{2m}v_{j}^{2n}=\tilde{\omega}^{4m},  \label{cct}
\end{equation}%
for the quantities $v_{j}=(\lambda _{j}^{4})^{\gamma /2}$. This is already
enough to prove that the number of degrees of freedom is finite, since (\ref%
{cct}) admits a finite number of solutions.

The right solutions $\lambda _{j}$ are those that solve (\ref{nleq}) as
well. In particular, the integral expressing $\ddot{x}(t)$ must be
convergent. Moreover, we know from appendix \ref{A} that the quantities $%
v_{j}$ must satisfy the inequality $|$Arg$[v_{j}]|\leqslant \pi \gamma /2$.

Now we prove that the acceptable $\lambda _{j}$ are the physical ones, which
are those that match the perturbative expansion of (\ref{nleq}): $\lambda
_{j}\simeq \pm \tilde{\omega}+\mathcal{O}(\tilde{\omega}^{2})$.

The first example we consider is the borderline case $m=1$, $n=1$, $\gamma
=1 $, which provides a way to treat the kinetic operator $\Box \left( 1+%
\sqrt{\Box ^{2}/M^{2}}\right) $. This is interesting because it contains the
\textquotedblleft absolute value\ $|\Box |$ of box\textquotedblright\ (in
the form$\sqrt{\Box ^{2}}$). We have $v_{j}=\sqrt{\lambda _{j}^{4}}$, so the
inequality $|$Arg$[(z^{4})^{\gamma /2}]|\leqslant \pi \gamma /2$ implies Re$%
[v_{j}]\geqslant $ $0$.

The $v_{j}$ equation (\ref{cct}) has a positive solution, which is%
\begin{equation*}
\nu^{\ast }=\frac{1}{2}\left( \sqrt{1+4\tilde{\omega}^{2}}-1\right) ,
\end{equation*}%
and three solutions with negative real parts, which must be discarded.
Inverting $\nu^{\ast }=\sqrt{\lambda _{j}^{4}}$ by means of (\ref{rz4}), we
find the four results $\lambda _{j}=(1,-i,i,-1)\sqrt{\nu ^{\ast }}$. When $%
\nu ^{\ast }\geqslant 1$ only the first and last options are acceptable
since the other two make the convolution of (\ref{nleq}) divergent, by (\ref%
{asy2}). However, when $\nu ^{\ast }<1$ all four make the convolution
convergent, which is a source of concern since the physical solutions are
only two.

This caveat was explained in ref. \cite{Calcagni}, where was shown that it
is not convenient to impose the fakeon prescription for special values of
the parameters (such as $\gamma =1$ here). Rather, the situation of interest
must be reached as a limit from a generic deformation. This means that we
need to work with an arbitrary, fractional $\gamma =m/n$, show that the
degrees of freedom are the physical ones there (which we do below) and take
the limit $\gamma \rightarrow 1^{-}$ at the end. In so doing, it becomes
visible that $\gamma =1$ only inherits the physical degrees of freedom.

The second example we consider is the case $m=1$, $n=2$, $\gamma =1/2$. Now, 
$v_{j}=(\lambda _{j}^{4})^{1/4}$ and the $v_{j}$ equation 
\begin{equation*}
\left( 1+v_{j}\right) ^{2}v_{j}^{4}=\tilde{\omega}^{4}
\end{equation*}%
has a unique acceptable solution, which is 
\begin{equation*}
\nu ^{\ast }=\frac{1}{3}\left( a+\frac{1}{a}-1\right) ,\qquad
a=2^{-1/3}\left( 3\tilde{\omega}\sqrt{81\tilde{\omega}^{2}-12}+18\tilde{%
\omega}-2\right) ^{1/3},
\end{equation*}%
since the other five have $|$Arg$[v_{j}]|>\pi /4$. Note that $\nu ^{\ast
} $ is real. Inverting (\ref{gz4}),\ we find $\lambda _{j}=(1,-i,i,-1)\nu
^{\ast }$, but only the first and last ones are acceptable, since by (\ref%
{asy}) the other two make the convolution of (\ref{nleq}) divergent.

The third example we treat is the case $m=1$, $n=4$, $\gamma =1/4$. Now $%
v_{j}=(\lambda _{j}^{4})^{1/8}$, and the $v_{j}$ equation 
\begin{equation*}
\left( 1+v_{j}\right) ^{2}v_{j}^{8}=\tilde{\omega}^{4},
\end{equation*}%
is solved by a unique real solution $\nu ^{\ast }$. The other seven
solutions violate the condition $|$Arg$[(z^{4})^{\gamma /2}]|\leqslant \pi
\gamma /2$. Inverting (\ref{gz4}),\ we find $\lambda _{j}=(1,-i,i,-1)(\nu
^{\ast })^{2}$, and again only $\lambda _{j}=\pm (\nu ^{\ast })^{2}$ are
acceptable, since the others make the convolution of (\ref{nleq}) divergent,
by (\ref{asy}).

Now we treat the general case for $\tilde{\omega}$ small. The solutions of (%
\ref{cct}) read%
\begin{equation*}
v_{j}\simeq -1+\cdots ,\qquad v_{j}\simeq \tilde{\omega}^{2m/n}\mathrm{e}%
^{\pi iq/n}+\cdots ,
\end{equation*}%
where $q$ is an integer. The first class of solutions is not acceptable
because they have a negative real part. As far as the second class is
concerned, the condition $|$Arg$[v_{j}]|\leqslant \pi \gamma /2$ implies $%
|q|\leqslant m/2$.

Now we have to solve%
\begin{equation*}
(\lambda _{j}^{4})^{m/(2n)}\simeq \tilde{\omega}^{2m/n}\mathrm{e}^{\pi
iq/n}+\cdots .
\end{equation*}%
Raising by $2n/m$, the phase satisfies $|\pi q/n\cdot 2n/m|=|2\pi q/m|$ $%
\leqslant \pi $, and the fourth root gives%
\begin{equation*}
\lambda _{j}\simeq (1,i,-i,-1)\tilde{\omega}\mathrm{e}^{\pi iq/(2m)}+\cdots .
\end{equation*}%
All such solutions have nonvanishing imaginary parts and make the
convolution of (\ref{nleq}) divergent, except for%
\begin{equation*}
\lambda _{j}\simeq (1,-1)\tilde{\omega}+\cdots .
\end{equation*}%
Thus, only these two solutions are acceptable, at least perturbatively in $%
\tilde{\omega}$. They are the expected, physical ones.

To conclude, the nonlocal structure of the equations does not burden (\ref%
{nleq}) with the need to fix other initial conditions besides the physically
expected ones, which are associated with the pole $p^{2}=0$ of propagators
such as (\ref{propa}).

\section{Conclusions}

\label{concl}\setcounter{equation}{0}

Quantum field theories with fractional or continuous powers of the
d'Alembert operator represent a challenging arena to extend knowledge about
QFT beyond its standard enclosure.

A well-defined classical limit with a finite number of initial conditions is
not guaranteed, because the field equations are nonlocal and generically non
real or non-Hermitian. At the quantum level, on the other hand, one must pay
attention to perturbative unitarity. Fakeons provide a simple way to
formulate unitary fractional models with an appropriate classical limit and
the right set of degrees of freedom.

The simplest formulation is the \textquotedblleft direct\textquotedblright\
one. It amounts to continuing the correlation functions from Euclidean space
to Minkowski spacetime using the fakeon approach for the fractional part of
the power. However, this is not the only way to employ fakeons for
fractional theories. We have shown that infinitely many alternatives are
available, and they are all inequivalent, in the sense that they generate
different correlation functions. In most cases, this is visible already at
the tree level. We also demonstrated it in bubble diagrams.

Finally, we have coupled fractional theories to gauge fields and gravity,
preserving the Ward and Cutkosky identities in all formulations.

Fractional models may or may not be useful to describe the physics of
fundamental interactions. However, in many cases they may describe systems
approximately as effective field theories. Potential applications range from
statistical systems to critical and off-critical phenomena, and more
generally descriptions of systems where certain effects of the experimental
apparatus or the boundary conditions drive the physics off-shell.

\vskip 1truecm \noindent {\large \textbf{Acknowledgments}}

\vskip .5truecm

The author is grateful to F. Briscese and G. Calcagni for useful discussions.

\vskip 1.2truecm

{\textbf{\huge Appendices}} \renewcommand{\thesection}{\Alph{section}} %
\renewcommand{\theequation}{\thesection.\arabic{equation}} %
\setcounter{section}{0}

\section{Fractional powers}

\label{A}\setcounter{equation}{0}

In this appendix we give details on the functions $z^{\gamma }$ and $%
(z^{2})^{\gamma /2}$, $z\in \mathbb{C}$, used in this paper. The function $%
z^{\gamma }$ is defined by the standard cut on the negative real axis $%
z\leqslant 0$. Then it follows that $(z^{2})^{\gamma /2}$ has a cut in Re[$%
z]=0$, which divides the complex plane into two disjoint regions.

There are two options to define $z^{\gamma }$ on the real axis $z=x\in 
\mathbb{R}$: $(x+i\epsilon )^{\gamma }$ and $(x-i\epsilon )^{\gamma }$,
where $\epsilon \rightarrow 0^{+}$. Both have nonvanishing imaginary parts
for $x<0$, which makes the field equations (\ref{feq}) not acceptable.

On the other hand, we can understand the function $(z^{2})^{\gamma /2}$ as $%
(x+i\epsilon )^{\gamma /2}(x-i\epsilon )^{\gamma /2}$ for $z=x\in \mathbb{R}$%
. If we write this expression as $(x^{2}+\epsilon ^{2})^{\gamma /2}$, the
parameter $\epsilon $ becomes irrelevant, hence we obtain $(x^{2})^{\gamma
/2}$, which is real on the whole real axis. This makes the field equations (%
\ref{feq2}) acceptable (Model 2).

Writing $z=\rho \mathrm{e}^{i\theta }$ on the complex plane we have%
\begin{equation}
(z^{2})^{\gamma /2}=\left\{ 
\begin{tabular}{l}
$\rho ^{\gamma }\mathrm{e}^{i\theta \gamma }\qquad $for $-\frac{\pi }{2}%
<\theta <\frac{\pi }{2},$ \\ 
$\rho ^{\gamma }\mathrm{e}^{i(\theta -\pi )\gamma }\qquad $for $\frac{\pi }{2%
}<\theta \leqslant \pi ,$ \\ 
$\rho ^{\gamma }\mathrm{e}^{i(\theta +\pi )\gamma }\qquad $for $-\pi
\leqslant \theta <-\frac{\pi }{2}.$%
\end{tabular}%
\right.  \label{zg2}
\end{equation}%
For example, $\sqrt{z^{2}}=z$ for Re[$z]>0$ and $\sqrt{z^{2}}=-z$ for Re[$%
z]<0$. In all cases, Re$[\sqrt{z^{2}}]\geqslant 0$. Moreover, $|$Arg$%
[(z^{2})^{\gamma /2}]|\leqslant \pi \gamma /2$. In particular, Re$%
[(z^{2})^{\gamma /2}]\geqslant 0$ for every $\gamma \leqslant 1$.

Formula (\ref{zg2}) decomposes $(z^{2})^{\gamma /2}$ in terms of its modulus
and argument, assuming the principal branch for fractional powers. The
condition on the argument is a consequence of first squaring $z$ and then
taking the square root of $z^2$.

We also have%
\begin{equation}
\sqrt{z^{4}}=\left\{ 
\begin{tabular}{l}
$z^{2}\qquad $for $|\theta |<\frac{\pi }{4}$ and $\frac{3\pi }{4}<|\theta
|\leqslant \pi $, \\ 
$-z^{2}\qquad $for $\frac{\pi }{4}<|\theta |<\frac{3\pi }{4}$,%
\end{tabular}%
\right. \qquad \text{Re}[(z^{4})^{1/2}]\geqslant 0.  \label{rz4}
\end{equation}%
with cuts on the diagonals of the complex plane. It is straightforward to
obtain (with the same cuts) 
\begin{equation}
(z^{4})^{\gamma /2}=\left\{ 
\begin{tabular}{ll}
$\rho ^{2\gamma }\mathrm{e}^{2\theta \gamma i}$ & for $-\frac{\pi }{4}%
<\theta <\frac{\pi }{4}$ \\ 
$\rho ^{2\gamma }\mathrm{e}^{(2\theta -\pi )\gamma i}$ & for $\frac{\pi }{4}%
<\theta <\frac{3\pi }{4}$, \\ 
$\rho ^{2\gamma }\mathrm{e}^{2(\theta -\pi )\gamma i}$ & for $\frac{3\pi }{4}%
<\theta \leqslant \pi $, \\ 
$\rho ^{2\gamma }\mathrm{e}^{(2\theta +\pi )\gamma i}$ & for $-\frac{3\pi }{4%
}<\theta <-\frac{\pi }{4}$, \\ 
$\rho ^{2\gamma }\mathrm{e}^{2(\theta +\pi )\gamma i}$ & for $-\pi \leqslant
\theta <-\frac{3\pi }{4}$,%
\end{tabular}%
\right.  \label{gz4}
\end{equation}
As before, $|$Arg$[(z^{4})^{\gamma /2}]|\leqslant \pi \gamma /2$.

\section{Analytic continuation of the bubble diagram}

\label{B}\setcounter{equation}{0}

In this appendix, we derive the analytic continuation (\ref{addi}) more
explicitly. For Re$[z]>0$, the integral (\ref{inti}) evaluates to 
\begin{equation*}
f(z,s,\sigma )=\frac{i}{\sqrt{(s+\sigma +iz)^{2}-4iz\sigma }}\ln \frac{%
s-\sigma +iz+\sqrt{(s+\sigma +iz)^{2}-4iz\sigma }}{s-\sigma +iz-\sqrt{%
(s+\sigma +iz)^{2}-4iz\sigma }}+\text{c.c.},
\end{equation*}%
It is easy to check that the function $f(z,s,\sigma )$ is analytic when the
arguments of the square roots become real and negative. Instead, it has a
cut for imaginary $z$, where the logarithm tends to $-i\pi $ from Re$[z]>0$.
When the imaginary axis is crossed from Re$[z]>0$ to Re$[z]<0$, the analytic
continuation gives and extra contribution obtained by replacing the
logarithm with $2\pi i$. The result 
\begin{equation*}
f^{\text{an}}(z,s,\sigma )=-f(-z,s,\sigma )+\frac{2\pi }{\sqrt{(s+\sigma
+iz)^{2}-4iz\sigma }}+\frac{2\pi }{\sqrt{(s+\sigma -iz)^{2}+4iz\sigma }}
\end{equation*}%
for Re$[z]<0$, matches the one of (\ref{addi}), having used (\ref{sim}).

As far as the function (\ref{g}) is concerned, we find%
\begin{equation*}
g(z,s,\sigma )=\frac{i}{\sqrt{(s-\sigma +iz)^{2}+4iz\sigma }}\ln \frac{%
s+\sigma +iz+\sqrt{(s-\sigma +iz)^{2}+4iz\sigma }}{s+\sigma +iz-\sqrt{%
(s-\sigma +iz)^{2}+4iz\sigma }}+\text{c.c}.
\end{equation*}%
for Re[$z]>0$ and $g^{\text{an}}(z,s,\sigma )=-g(-z,s,\sigma )$ for Re[$z]<0$%
.

\end{document}